\documentclass[lettersize,journal]{IEEEtran}
\IEEEoverridecommandlockouts
\usepackage{amsmath,amsfonts}
\usepackage{algorithmic}
\usepackage{array}
\usepackage[caption=false,font=normalsize,labelfont=sf,textfont=sf]{subfig}
\usepackage{textcomp}
\usepackage{stfloats}
\usepackage{url}
\usepackage{verbatim}
\usepackage{graphicx}
\usepackage{xcolor}
\usepackage{cite}
\usepackage{graphicx}
\usepackage{authblk}
\usepackage{float}
\usepackage{threeparttable}
\newcommand{\insertWideFigure}[2]{
    \begin{figure*}[b]
        \vspace{-0.4cm}
        \centering
        \includegraphics[width=\textwidth]{#1.pdf}
    	\vspace{-0.7cm}
        \caption{\small #2}
        \label{fig:#1}
    \end{figure*}
}

\def\BibTeX{{\rm B\kern-.05em{\sc i\kern-.025em b}\kern-.08em
    T\kern-.1667em\lower.7ex\hbox{E}\kern-.125emX}}
\usepackage{balance}

\begin{document}
%\title{Dynamic Modelling of the Phase Change in Liquid Crystal Antennas for Switching Times Reduction}
%\title{Dynamic Modelling of Liquid Crystal Reflectarray Cells and its Application to Reducing Switching Times}
%\title{Dynamic Modelling of Liquid Crystal Reflectarray Cells and its Application to Overdriving Techniques}
%\title{Dynamic Modelling of Liquid Crystal Reflectarray Cells and its Application to Timely Driving Techniques}
%\title{Dynamic Modelling of Liquid Crystal Planar Multiresonant Cells and its Application to Reducing Transition Times Between Phase-Shifting States}
%\title{Dynamic Modelling of Planar Multiresonant Cells Based on Liquid Crystal and its Application to Reducing Reconfigurability Times}
%\title{Dynamic Modelling of Liquid Crystal Planar Multiresonant Cells and its Application to Reducing Reconfigurability Time Between Phase States}
%\title{Dynamic Modelling of Liquid Crystal Planar Multiresonant Cells and its Application to Reducing Reconfigurability Times}
\title{Dynamic Modelling of Liquid Crystal-Based Metasurfaces and its Application to Reducing Reconfigurability Times}
%\titleheader{This paper has been accepted at IEEE Transactions on Antennas and Propagation}

\begin{comment}
\author[*]{Robert Guirado}
\author[*]{Gerardo Perez-Palomino}
\author[*]{Marta Ferreras}
\author[*]{Eduardo Carrasco}
\author[$\dagger$]{Manuel Caño-García}

\affil[*]{IPTC, Universidad Polit\'{e}cnica de Madrid, Spain}
\affil[$\dagger$]{CEMDATIC, Universidad Polit\'{e}cnica de Madrid, Spain}%\thanks{This work was developed by...}}
\end{comment}
\author{Robert Guirado, Gerardo Perez-Palomino, Marta Ferreras, Eduardo Carrasco, and Manuel Caño-García
\thanks{This work was supported in part by the Spanish Ministry of Science and Innovation and the Spanish Agency for
Research within projects ENHANCE-5G (PID2020-114172RB-C22/AEI/10.13039/501100011033) and TSI-063000-2021-83, and by the European Union’s Horizon 2020 Framework Programme for Research and Innovation: ARIADNE project, under grant agreement no.871464. R. Guirado acknowledges the support of a fellowship from ”la Caixa” Foundation (ID 100010434). The fellowship code is LCF/BQ/DR21/11880029. M. Caño-García is grateful to Spanish government grant (BG20/00136).}
\thanks{R. Guirado, G. Perez-Palomino and E. Carrasco are with the Group of Applied Electromagnetics (GEA), ETSI Telecomunicación,
Universidad Politécnica de Madrid, E-28040 Madrid, Spain (e-mail:
r.guirado@upm.es).}%
\thanks{M. Ferreras is with the Microwave and Radar Group, ETSI Telecomunicación, Universidad Politécnica de Madrid, E-28040 Madrid, Spain.}%
\thanks{M. Caño-García is with the CEMDATIC, ETSI Telecomunicación, Universidad Politécnica de Madrid, E-28040 Madrid, Spain.}}%

\maketitle

\makeatletter
\def\ps@IEEEtitlepagestyle{
  \def\@oddfoot{\mycopyrightnotice}
  \def\@evenfoot{}
}
\def\mycopyrightnotice{
  {\tiny
  \begin{minipage}{\textwidth}
  \centering
  \copyright 2022 IEEE. Personal use of this material is permitted. Permission from IEEE must be obtained for all other uses, in any current or future media, including reprinting/republishing this material for advertising or promotional purposes, creating new collective works, for resale or redistribution to servers or lists, or reuse of any copyrighted component of this work in other works.
  \end{minipage}
  }
}
%\IEEEpubid{\begin{tabular}[t]{@{}l@{}} \vspace{2cm}{ \tiny \copyright 2022 IEEE. Personal use of this material is permitted. Permission from IEEE must be obtained for all other uses, in any current or future media, including reprinting/republishing this material for advertising or promotional purposes, creating}\\{\tiny new collective works, for resale or redistribution to servers or lists, or reuse of any copyrighted component of this work in other works.}\end{tabular}}
%\IEEEpubid{\begin{minipage}{\textwidth}\ \\[10pt] \centering
%  \vspace{-0.5cm}
%  {\tiny \copyright 2022 IEEE. Personal use of this material is permitted. Permission from IEEE must be obtained for all other uses, in any current or future media, including reprinting/republishing this material for advertising or promotional purposes, creating\\ new collective works, for resale or redistribution to servers or lists, or reuse of any copyrighted component of this work in other works.}
  %See http://www.ieee.org/publications standards/publications/rights/index.html for more information.
  %20xx IEEE. Personal use of this material is permitted. Permission from IEEE must be obtained for all other uses, in any current or future media, including reprinting/republishing this material for advertising or promotional purposes, creating new collective works, for resale or redistribution to servers or lists, or reuse of any copyrighted component of this work in other works.
%\end{minipage}} 

\begin{abstract}
This paper describes and validates for the first time the dynamic modelling of Liquid Crystal (LC)-based planar multi-resonant cells, as well as its use as bias signals synthesis tool to improve their reconfigurability time. The dynamic LC director equation is solved in the longitudinal direction through the finite elements method, which provides the $z$- and time-dependent inhomogeneous permittivity tensor used in an electromagnetic simulator to evaluate the cells behaviour. The proposed model has been experimentally validated using reflective cells for phase control (reflectarray) and measuring the transient phase, both in excitation and relaxation regimes. It is shown how a very reduced number of stratified layers are needed to model the material inhomogeneity, and that even an homogeneous effective tensor can be used in most of the cases, which allows a model simplification suitable for design procedures without losing accuracy. Consequently, a novel bias signal design tool is proposed to significantly reduce the transition times of LC cells, and hence, of electrically large antennas composed of them. These tools, similar to those used in optical displays, are experimentally validated for the first time at mm- and sub-mm wave frequencies in this work, obtaining an improvement of orders of magnitude.

\end{abstract}

\markboth{Journal of XXX,~Vol.~XX, No.~XX, September~XXX}%
{}

\vspace{-0.4cm}
\section{Introduction}

Liquid Crystal (LC) technologies are starting to be studied at mm-wave bands in order to develop tunable devices that work properly at those frequency ranges. Because of its birefringence, by applying a low-frequency electric field to a cavity containing nematic LC, its electromagnetic properties can be varied and therefore the device response changes \cite{fundLC}. This phenomenon has been widely used in optics to develop LC displays and other devices such as spatial phase modulators \cite{spm}, but its potential use at mm-wave frequencies has only started to flourish \cite{employing}. 

This varying behaviour is related to a continuous change on the electric permittivity, which can be leveraged to vary the resonant frequency or beam of an antenna \cite{lcantenna, lcantenna2, lcantenna3, toso}, to sweep the shift in phase shifters \cite{LCphaseshifter, fast, phaseshift, reconf} or to tune the different elements in a reflectarray antenna \cite{lcreflectarray, reflectarray2,reflectarray3,folded,svhum}, to name a few. Very recently, the use of nematic LC as the key phase-shifting element of the upcoming Reconfigurable Intelligent Surfaces (RIS), planar devices able to manipulate electromagnetic propagation, has been proposed \cite{lcIRS,metasurf,lcmetasurface}, as it is one of the few tunable technologies with moderate cost capable of keeping the pace of the high frequencies ($>$100GHz) expected in future network generations beyond 5G. %(e.g. 6G).  
Moreover, since the LC fills an entire cavity and its behaviour can be locally modified, a pixel-wise active element implementation is avoided. This, together with the fact that these manufacturing procedures are widely common in optics, especially when developing electrically large planar devices with thousands of cells, make of this technology a very attractive solution for developing RIS panels. Alternative solutions to switch a beam in a metasurface, such as mechanical steering \cite{mech} or unit cells based on varactor diodes \cite{varactor}, although being commercially available are either of much higher cost or exhibit frequency limitations, as Table \ref{tab:techs} shows.

Nevertheless, the relatively large losses and the slow switching times between states are the main weaknesses of such LC-based devices, as detailed in Table \ref{tab:techs}. Even though LC manufacturers are starting to develop novel composites specifically designed to present low losses at microwave and mm-wave frequencies \cite{employing}, current mm-wave LC devices provide reconfigurability times in the order of the seconds. However, to be fully implantable in future ultra-reliable low-latency communication networks, improving these times until they are at least comparable to the channel coherence time is of utmost importance due to the stringent dynamic requirements of upcoming communication protocols. 
\markboth{This paper has been accepted at IEEE Transactions on Antennas and Propagation. DOI: 10.1109/TAP.2022.3209734}%
{}
A number of strategies have been reported in optics to improve transition times, such as the use of polymerizable compounds \cite{pnlc,pnlc2} or dual-frequency LC \cite{dflc,dflc2,dflc3}, especially for decay time. Another strategy to reduce transition times is to employ sophisticated excitation signals by leveraging LC dynamics \cite{overdrive, overdrive_optics,overdrivelow}. In order to understand and completely control its dynamic behaviour during a state transition, which will ultimately impact on the switching time, its accurate temporal modelling is essential. In spite of this, few works tackle the temporal aspects of LC between arbitrary states \cite{smallangle,correlations,templc,comprehensive}, and all of them focus in the optical regime and not in RF. Moreover, given the challenge of achieving a proper phase shift range in mm-wave bands, multi-resonant cells (i.e. including several resonant elements in a single-band cell) must be used \cite{lcreflectarray,gerardorefl}. This makes modelling much more complex \cite{3dmodel} since resonators create phase shifts that can not be modelled with a medium constant, and they are also used to locally bias the LC. In \cite{accurate} an accurate LC modelling at mm-wave frequencies was proposed for cells in which it is necessary to include resonant elements. Using that model, it was shown that inhomogeneity and anisotropy can be considered only in the longitudinal direction of the cell, and the minimum number of layers in which the media should be stratified to obtain a precise phase prediction was also found. However, that model only considers a static LC regime, that is, when enough time has passed after an external excitation so that molecules lie in a stationary state after rotating. In the nematic LC characteristic equation, this translates in neglecting time-dependent terms. Consequently, there is no previous research in accurately characterizing the dynamics of LC-based mm-wave devices to reduce switching times.

In this paper, we accurately model for the first time the LC dynamics between transition states (Fréedericksz transition) in RF past the known approximations, in order to obtain a temporal design control capability of the LC, representing a contribution to the model beyond the one reported in \cite{accurate}. First, we solve the LC dynamics equation applied to multi-resonant cells, analyzing its convergence as a function of the number of layers. This is done to investigate whether is it possible to compress the LC inhomogeneous molecular orientation in a single layer of an effective permittivity tensor, computed with the average molecule tilt across the cavity, with the aim of enabling a much more efficient electromagnetic cavity evaluation. The validation of the proposed model is carried out experimentally through different state transitions.
As another novelty of this work given its applicability, the proposed and validated modelling is used to introduce an efficient design procedure of the LC bias signal, similar to the overdrive technique used in optical panels, capable of diminishing the switching time (especially rise transitions) of planar multi-resonant cells for phase control by a factor of 100X. Finally, both the model and the overdrive excitation technique are validated by comparing simulation and experimental data, obtained from reflective multi-resonant cells fed by a plane wave (reflectarray, metasurface or RIS), although the model is extendable to transmissive structures (transmitarrays) or other planar structures capable of controlling other parameters than phase. In the paper, we discuss the reconfigurability time improvement as a function of the used biasing signals, showing that the proposed method mitigates one of the most important challenges this technology must overcome to be widely used.

The model results are validated at the cell (pixel) level for two different frequencies (97 GHz and 102 GHz), which in turn facilitates a tool for the analysis and synthesis of control signals at an arbitrary frequency and per each cell of the whole antenna, given that in a complete surface a plane wave with a different incident angle will arrive to each pixel. Therefore, this tool allows to synthesize overdriving control signals without depending on experimental measurements for each cell and angle of incidence in the array.
\markboth{Journal of XXX,~Vol.~XX, No.~XX, September~XXX}%
{}

The paper is organized as follows. Section II introduces the dynamic modelling of LC multi-resonant cells and discusses analytical solutions to motivate the use of overdriving to reduce transition times. In Section III, the dynamic LC differential equation is solved exactly by combining a finite elements method (COMSOL) and an full-wave electromagnetic analysis tool (CST), the phase convergence of the cell is analyzed and the model is validated. Then, in Section IV, the overdrive design tool is presented and experimentally tested, and Section V concludes.

\begin{table}[h]
    \centering
    \vspace{-0.2cm}
    \caption{Beam switching technologies comparison}
    \begin{tabular}{|c|c|c|c|c|c|}
        \hline
        \textbf{Strategy} & \textbf{Speed} & \textbf{Freq. range} & \textbf{Cost} & \textbf{Efficiency} & \textbf{Energy} \\
        \hline
        \textbf{Mechanical} & $s$ & Wide & High & Low & High \\
        \hline
        \textbf{Varactor} & $\mu s$ & Limited & Medium & High & Medium\\
        \hline
        \textbf{LC} & $s$ & Wide & Low & Low & Very low \\
        \hline
    \end{tabular}
    \vspace{0.1cm}
    \label{tab:techs}
    \vspace{-0.4cm}
\end{table}

\section{Dynamic Modelling of LC resonant cells}

\subsection{Permittivity Tensor Calculation}
\label{sec:Permittivity Tensor Calculation}
\begin{figure}[!htb]
    \centering
    \vspace{-0.2cm}
    \includegraphics[width=1\columnwidth]{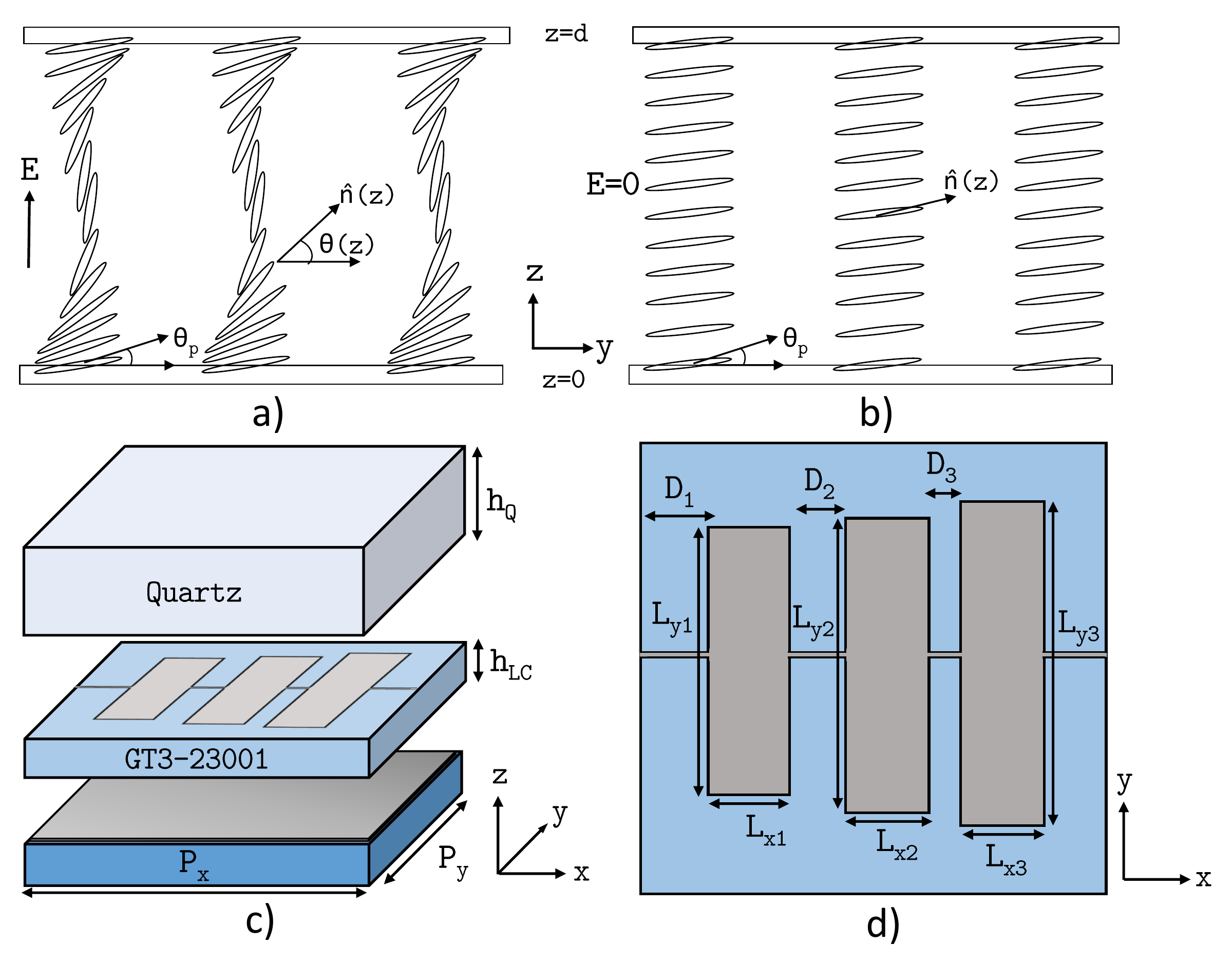}
    \vspace{-0.5cm}
    \caption{a) LC cavity with applied electric field b) LC cavity without excitation c) Layered view of the reflectarray unit cell d) Top-view of the reflectarray unit cell. Dimensions (mm): $D_1 = 0.171$, $D_2 = 0.096$, $D_3 = 0.042$, $L_{y1} = 0.707$, $L_{y2} = 0.748$, $L_{y3} = 0.792$, $L_{x1} = 0.2$, $L_{x2} = 0.211$, $L_{x3} = 0.2$, $P_{x} = 1.145$, $P_y = 1.093$, $h_Q = 0.55$, $h_{LC} = 0.075$.}  %\textcolor{red}{modificar?}}
    \label{fig:cavity}
    \vspace{-0.2cm}
\end{figure}
LC-based mm-wave devices leverage the tuning capability of this material, typically used as a substrate, to become dynamically reconfigurable. A typical LC cavity is shown in Fig. \ref{fig:cavity}a and \ref{fig:cavity}b with and without external excitation respectively, where the $z=0$ and $z=d$ planes are assumed to be indefinite electrodes. In this case, the rod-like molecules only present inhomogeneity along $z$ when biased. In the nematic state, due to the small degree of positional order of nematic LCs, a low-frequency (AC) electric field applied across the cavity will rotate its molecules. This, together with the LC anisotropy (i.e. large orientational order), allows the permittivity to be varied, as shown in Equation (\ref{eq:eptensor}). 

\begin{equation}
    \label{eq:eptensor}
    \overline{\overline{\varepsilon_r}}(\overline{r},t) = \varepsilon_{r\perp} \overline{\overline{I}} + \Delta\varepsilon_r \overline{\overline{N}}(\overline{r},t), 
\end{equation}
where $\overline{\overline{I}}$ is the 3x3 identity matrix and $\overline{\overline{N}}(\overline{r},t) = \hat{n}(\overline{r},t) \otimes \hat{n}(\overline{r},t)$, being $\hat{n}(\overline{r},t)$ the macroscopic vector that defines the local orientation of the LC molecules at a certain point and time. This way, if an external electric field changes 
$\overline{\overline{N}}$, the macroscopic permittivity will be tuned.
Given that the substrate under use is both inhomogeneous and anisotropic, its dielectric permittivity has to be expressed as a tensor, $\overline{\overline{\varepsilon_r}}$. The dielectric anisotropy of the material is defined as $\Delta\varepsilon_r = \varepsilon_{r||}-\varepsilon_{r\perp}$, being $\varepsilon_{r||}$ and $\varepsilon_{r\perp}$ the parallel and perpendicular dielectric constants with respect to $\hat{n}$, which respectively relate the parallel and perpendicular components of the RF electric field to the electric displacement field. A concurrent problem in LC-based mm-wave designs is that the birefringence of LC cells has been typically underestimated and modelled with a scalar permittivity value, ranging $\varepsilon_{\perp}>\varepsilon_r>\varepsilon_{||}$, instead of using its tensor form. This oversimplifies the problem by transforming it into an isotropic scenario, as by working with only the scalar permittivity the results are geometry-dependent and do not generalize, being insufficient for an accurate modelling.

%\textcolor{red}{explain eq. diferencial? functional..}

\begin{figure*}[b]
    \vspace{-0.4cm}
    \centering
    \includegraphics[width=1\textwidth]{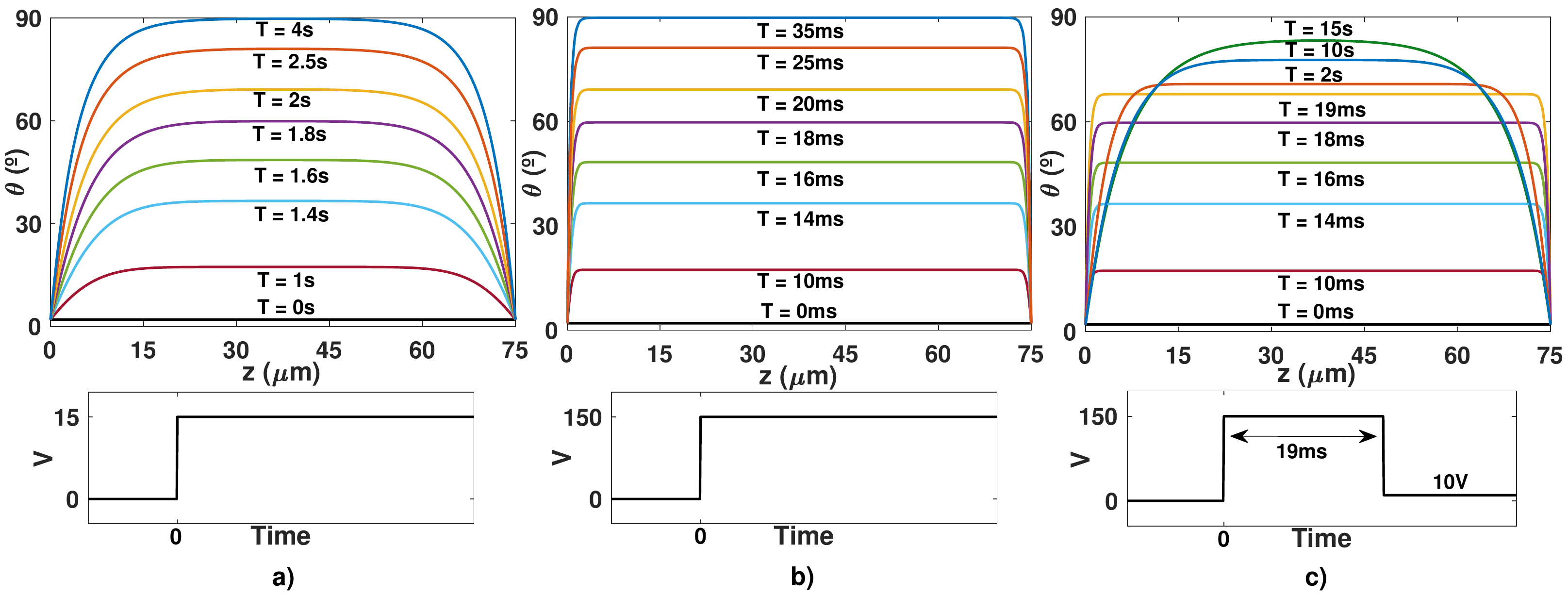}
    \vspace{-0.8cm}
    \caption{Tilt angle dynamics in a $75\mu m$ thick cavity filled with GT3-23001 LC. a) Step transition from 0V to 15V b) Step transition from 0V to 150V c) Transition from 150V to 10V after 19ms.}
    \label{fig:transition_up}
    \vspace{-0.2cm}
\end{figure*}

Both in reflectarray antennas and RIS, this dielectric anisotropy is used to perform a pixel-wise phase shift across an impinging electromagnetic wave. By applying a pre-computed bias voltage to each cell, the reflection coefficient phase is locally modified, thus obtaining a desired global phase distribution at the output, which will dictate the direction of the reflected wave. As shown in Fig. \ref{fig:cavity}c and \ref{fig:cavity}d, a unit cell of such reflectarray antennas consists of: a set of electrodes (typically dipoles, connected to the same potential within the cell) printed below a superstrate (e.g. quartz), which will act as a top plate of the LC cavity; the LC layer itself; a bottom conductive plate; and a substrate to support the structure. When carefully designed, the role of the dipoles is to produce an appropriate resonance in the reflection coefficient in RF, which will ultimately create a phase shift depending on the biasing. 
However, besides their role in RF, the dipoles also have a function in AC, since they are typically used to polarize the LC with the low-frequency electric field. This makes the tensorial permittivity to actually be inhomogeneous also in the transverse directions ($x$ and $y$), although in \cite{accurate} it was shown in the static case how its effects in the reflected phase are negligible as compared to the inhomogeneity and anisotropy effects in the longitudinal direction. In this paper, it will be shown how these effects also have little impact when modelling LC dynamics. Notwithstanding, in \cite{anibal,3dmodel} a more rigorous LC modelling is introduced for complex structures in which these effects are not negligible.

It is worth noting that in literature there exist three general strategies to perform 2D addressing of the cells, in order to apply the proper voltage (and therefore, to obtain the proper phase) to each antenna pixel: direct, active and passive addressing \cite{fundLC}. In the cases of active or passive addressing, the sequential row sweep implies the need of synthesizing voltage sequences which must be properly computed, requiring a tool to analyse and synthesize control signals, like the one introduced in this paper. In the direct addressing case, the proposed model and tool are especially useful to reduce transition times between states, enabling overdrive techniques.

The complex task of synthesizing control sequences could alternatively be done through measurement data instead of simulations. However, as long as sufficiently accurate models are used, the latter greatly simplifies the process given that the LC dynamic effects depend on the cell position within the array (different incidence angle), the dimensions of the resonators in each cell (cells could have different dimensions in each pixel), the operating frequency and the LC properties.

\vspace{-0.2cm}
\subsection{Dynamic Director Calculation}
Assuming that the applied AC electric field is homogeneous across the cell, which is feasible if the effects of transverse inhomogeneity are negligible, the dynamic behaviour of the LC under a such external excitation is described by the Ericksen–Leslie equation \cite{ericksen,leslie, fundLC}:
\vspace{-0.2cm}
\begin{multline}
    \label{eq:ericksenleslie1}
    (k_{11}cos^2\theta + k_{33}sin^2\theta) \frac{\partial^2\theta}{\partial z^2} + (k_{33}-k_{11}) \cdot sin\theta \cdot cos\theta(\frac{\partial\theta}{\partial z})^2 + \\(\alpha_2 sin^2\theta - \alpha_3cos^2\theta)\frac{\partial v}{\partial z} + \varepsilon_0 E^2 \Delta\varepsilon_q\cdot sin\theta \cdot cos\theta\\ = \gamma_1\frac{\partial\theta}{\partial t} + I\frac{\partial^2\theta}{\partial t^2},
\end{multline}
where $k_{11}$ and $k_{33}$ are the splay and bend elastic constants, $\alpha_i$ is the Leslie viscosity coefficient, $v$ is the flow velocity, $E = V_q/d$ is the applied quasi-static electric field, $\Delta\varepsilon_q$ is the low-frequency dielectric anisotropy, $\theta$ is the tilt angle of the director, $\gamma_1$ is the rotational viscosity and $I$ represents the inertia. Two modifications can be safely made to the previous equation. First, the inertia term is typically neglected as it has a very small weight \cite{chand}. Second, the Leslie coefficients terms can be disregarded, as they also play a minor role and obtaining them requires experimental measurements \cite{lesliecoef,fundLC}. Their impact will be seen later on in relation to backflow effects. Therefore, the previous equation reduces to \cite{smallangle,pretilt}:
\begin{multline}
    \label{eq:ericksenleslie2}
    (k_{11}cos^2\theta + k_{33}sin^2\theta) \frac{\partial^2\theta}{\partial z^2} + (k_{33}-k_{11}) \cdot \\ sin\theta \cdot cos\theta(\frac{\partial\theta}{\partial z})^2 + \varepsilon_0 E^2 \Delta\varepsilon_q\cdot sin\theta \cdot cos\theta = \gamma_1\frac{\partial\theta}{\partial t},
\end{multline}
which allows the time-varying director to be found.

%\textcolor{red}{Explain Vth}
In order to solve Equation (\ref{eq:ericksenleslie2}), boundary conditions for $\theta_{(z=0)}$ and $\theta_{(z=d)}$ must be applied. In the ideal case, $\theta_{(z=0)}=\theta_{(z=d)}=\theta_p=0$ (zero pre-tilt), which makes Equation (\ref{eq:ericksenleslie2}) stable from a certain threshold voltage, $V_{th}$. Then, the LC molecules will not start reorienting until this threshold voltage is reached. Specifically, $V_{th}$ can be computed as:

\begin{equation}
    \label{eq:vth}
    V_{th} = \pi\sqrt{\frac{k_{11}}{\Delta\varepsilon_q}}
\end{equation}

In a real implementation of a LC cell, the inner-most surface of the two conductive plates contains a rubbed polyimide orienting layer, which forces the pre-tilt boundary condition $\theta_p\neq0$ at $z=0$ and $z=d$. This ensures that the molecules are correctly oriented also in absence of excitation, as can be seen in Fig. \ref{fig:cavity}b. Under these conditions, $V_{th}$ is not a strict value anymore, as molecules are able rotate even below this threshold. However, it can still be used as a reference to identify a voltage point in which the LC starts reacting more energetically, since below it the molecular orientation is weak.

Equation (\ref{eq:ericksenleslie2}) entails the dynamic behaviour of the LC under an excitation, but since it is too complex to be solved analytically, certain assumptions are typically made in order to obtain approximate analytical solutions. Specifically, it is usually assumed that $k_{11} = k_{33}$, and that the LC is excited with a low voltage source. This allows to model the $\theta(z)$ curve with a sinusoidal function, so that $sin(\theta)\sim\theta$, which greatly simplifies the previous expression, resulting in Equation (\ref{eq:simplifiedLes}).

\begin{equation}
\label{eq:simplifiedLes}
    k_{11}\frac{\partial^2\theta}{\partial z^2} + \varepsilon_0 E^2 \Delta\varepsilon_q\theta = \gamma_1\frac{\partial\theta}{\partial t}
\end{equation}

Then, the solution for the tilt angle along $z$ takes, in time, an exponential form with the following decay and rise time constants:
\begin{equation}
    \tau_d = \gamma_1 \frac{d^2}{k_{11}\pi^2},
    \label{eq:decay}
\end{equation}
\begin{equation}
    \tau_r = \frac{\tau_d}{\left|\left(\frac{V}{V_{th}}\right)^2-1\right|},
    \label{eq:rise}
\end{equation}

However, these approximations oversimplify the dynamics problem as (i) they assume a pretilt angle equal to zero, which is not realistic \cite{correlations}; (ii) they are only valid for small voltage excitation where the tilt can be approximated with a sine, which is a rough estimation; (iii) they assume that the driving voltage is little above $V_{th}$, while we will later show that voltages much greater than that are needed in order to accelerate the LC response. Although more elaborated expressions have been introduced in \cite{pretilt} to include pretilt effects, small angle approximations and single elastic constants ($k = k_{11} = k_{33}$ or $k = (k_{11}+k_{33})/2$) are still assumed.
In \cite{accurate}, the static voltage dependence of LC is studied and accurately predicted but the dynamics are not tackled. In \cite{fast}, the LC temporal behaviour is experimentally measured for different commercially available materials in $4\mu$m thick cells, but a model is not provided.

\begin{figure}[!htb]
    \centering
    \includegraphics[width=1\columnwidth, height=4.5cm]{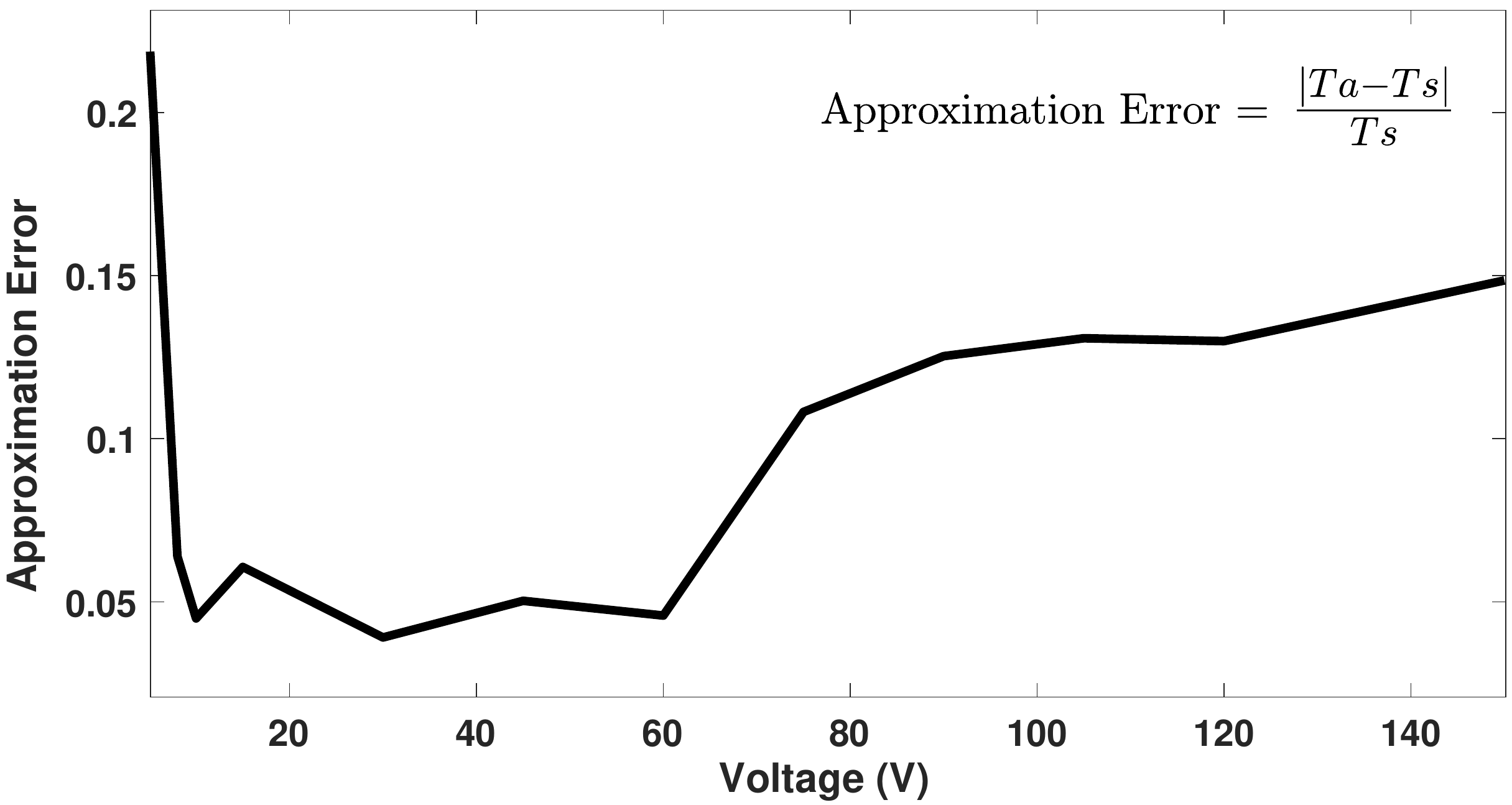}
    \vspace{-0.7cm}
    \caption{10\%-90\% rise time approximation error compared to the solver solution in a $75\mu m$ thick cavity. $Ta$ refers to the approximated time from the closed expressions and $Ts$ refers to the simulated time from Eq. (\ref{eq:ericksenleslie2}). It can be observed how the closed expressions fail for large voltages, as the sinusoidal approximation becomes invalid, as well as the Vth effect.}
    \label{fig:approx_error}
    \vspace{-0.3cm}
\end{figure}

Moreover, in optics, the phase change that occurs from this tilt dynamics can be well modelled, and the optical intensity change can be well predicted, since for such a short wavelength the LC is simply a medium in which several $\pi$-times phase changes happen \cite{correlations}. %There, $delta(t) = sin^2()$... is known.
However, in mm-wave devices, extra resonant elements are needed to enlarge the phase range up to a few $\pi$-times, which makes the relationship between the tilt angle dynamics and phase changes much more complex to model. When including resonant elements such as printed dipoles, the electric field in the cavity contains significant components in all directions and anisotropy can not be overlooked. That is, since the phase change is generated through printed metallizations in the superstrate, the cavity can not be modelled anymore with a medium constant. Furthermore, in optics, the cavities are typically much narrower ($\sim10um$) than in microwaves ($\sim100um$), being such thick cavities quite unexplored and not modelled. Therefore, it is necessary to solve Equation (\ref{eq:ericksenleslie2}) for resonant cells without any of these limitations, so that an appropriate model is achieved considering the number of layers and the inhomogeneity to consider, and to be validated with measurement data.

\begin{comment}
\begin{figure}[!htb]
    \centering
    \includegraphics[width=0.7\columnwidth]{images/diagrama.pdf}
    \vspace{-0.6cm}
    \caption{Block diagram.}
    \label{fig:diagrama}
    \vspace{-0.6cm}
\end{figure}
\end{comment}

\begin{figure}[!htb]
    \centering
    \vspace{-0.2cm}
    \includegraphics[width=1\columnwidth]{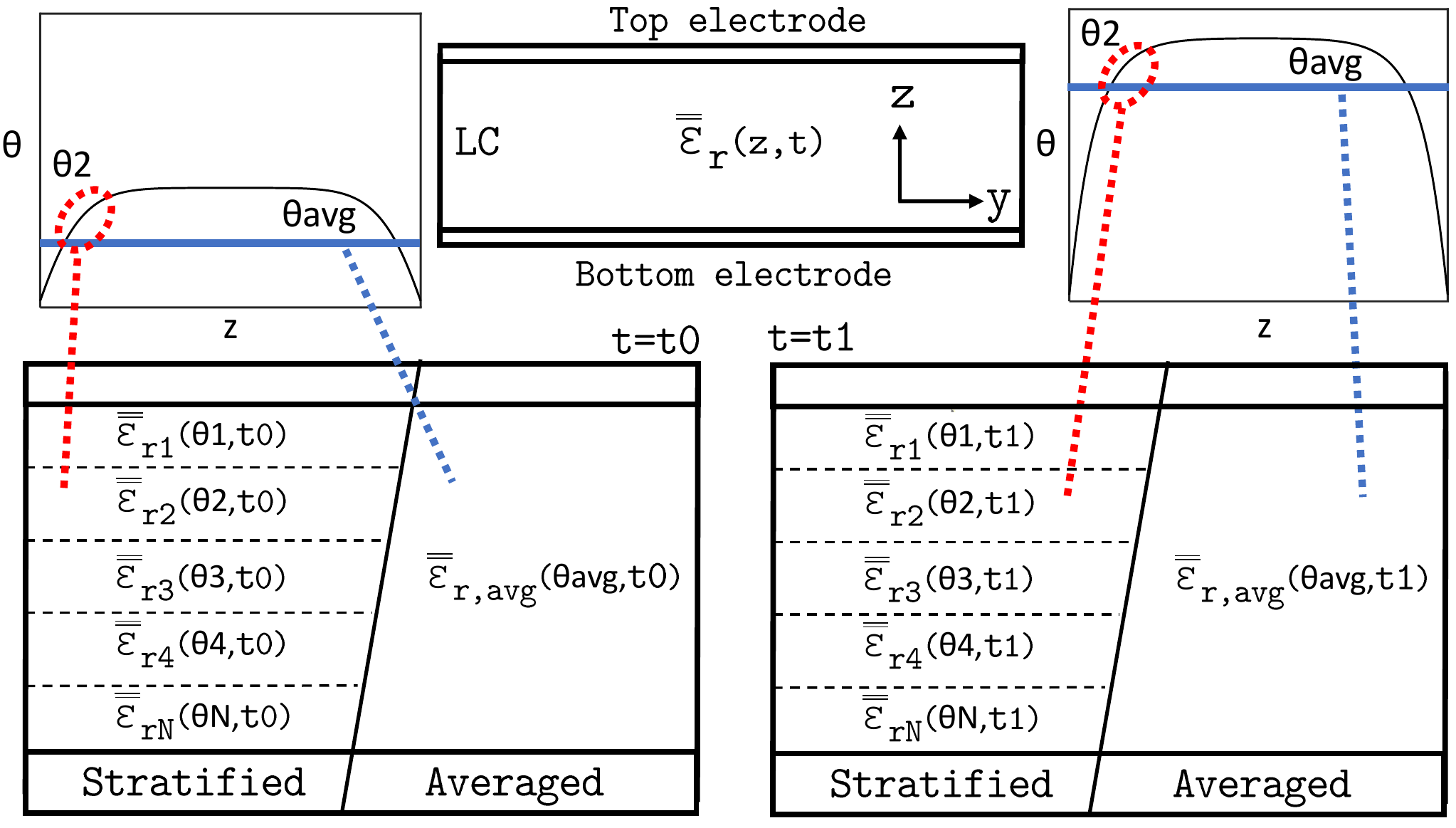}
    \vspace{-0.6cm}
    \caption{Stratified and averaged LC cavity dynamic modelling strategies. In the stratified strategy, $\overline{\overline{\varepsilon}}_{ri}(\theta_i,t)$ is computed with Eq. (\ref{eq:eptensor}) and considering as $\theta_i$ the average tilt within the layer $i$. In the averaged strategy, $\overline{\overline{\varepsilon}}_{r,avg}(\theta_{avg},t)$ is computed considering as $\theta_i$ the average tilt across the entire cavity.}%$\overline{\overline{\varepsilon_r}}(z_{i..j})$ is computed with Eq. (\ref{eq:eptensor}) and considering as $\theta$ the average tilt angle between $z_i$ and $z_j$.}
    \label{fig:stratified}
    \vspace{-0.4cm}
\end{figure}

\section{Model Results and Experimental Validation}
\label{sec:Model Results and Experimental Validation}

In order to accurately model the tilt angle dynamics of the LC when an excitation change occurs, we computationally solve Equation (\ref{eq:ericksenleslie2}) along $z$ and $t$ using finite elements method in COMSOL Multiphysics \cite{comsol}. This avoids several error sources as we specifically consider the pretilt effects and a more complete set of elastic constants, obtaining a more precise data for the $\theta(z)$ curves at any timestamp and LC driving voltage. Moreover, this allows to model any kind of excitation beyond step-like functions, although if the excitation signal has a frequency high enough so that its period is much smaller than the relaxation time of the LC, it can be substituted by its root mean square (RMS) value in Equation (\ref{eq:ericksenleslie2}).

\insertWideFigure{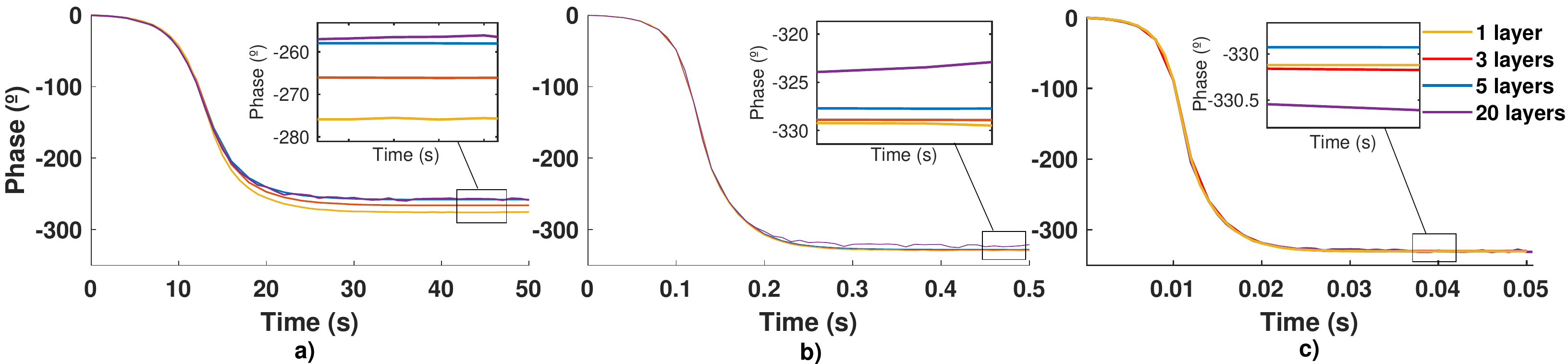}{Stratified simulation convergence study at 97 GHz a) from 0V to 5V b) from 0V to 45V c) from 0V to 150V.}

Fig. \ref{fig:transition_up} shows the molecules tilt angle as a function of $z$ for different timestamps in three different excitation scenarios, in a $75\mu m$ thick cavity filled with GT3-23001 LC. It can be observed how large voltage excitations make transitions faster and much more homogeneous tilts. By using the proposed tool, Fig. \ref{fig:approx_error} shows the error on the computed rise time (10\% to 90\%) when the sine approximation and Equations (\ref{eq:decay}) and (\ref{eq:rise}) are used with respect to the real solution of Equation (\ref{eq:ericksenleslie2}) for the same cavity. As expected, it is necessary to model the cell rigorously especially in the high voltage regime, given that errors larger than 10\% could be made in predicting the LC temporal behaviour, which in turn induces large errors in the cell RF reflected phase as a function of time.

\begin{figure*}[h!]
    \centering
    \includegraphics[width=1\textwidth]{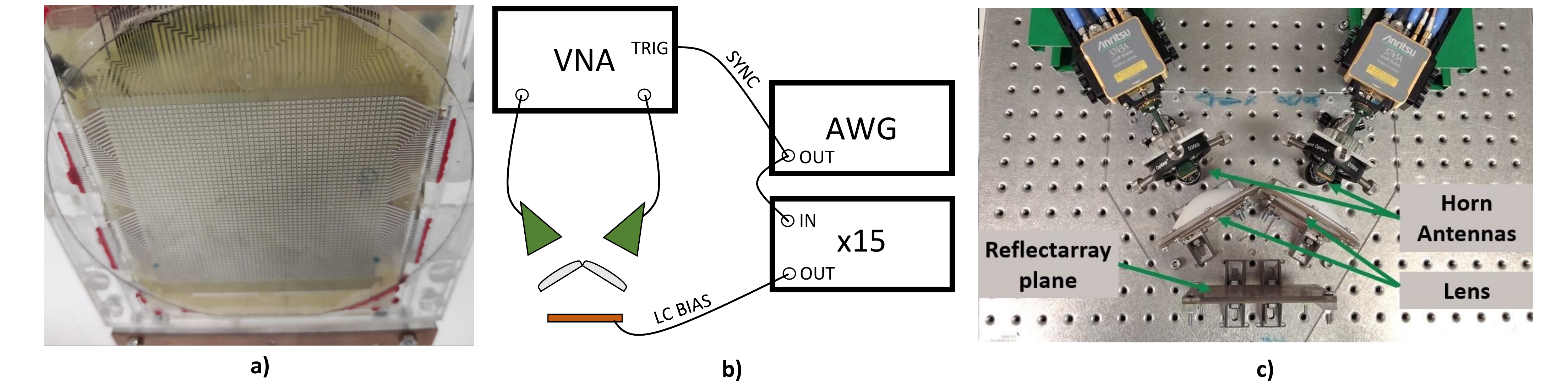}
    \vspace{-0.7cm}
    \caption{Measurement setup. a) Reflectarray picture b) Block diagram of the setup c) Quasi-optical bench picture}
    \label{fig:setup}
    \vspace{-0.5cm}
\end{figure*}

\begin{figure}[!htb]
    \centering
    \includegraphics[width=1\columnwidth, height=5.5cm]{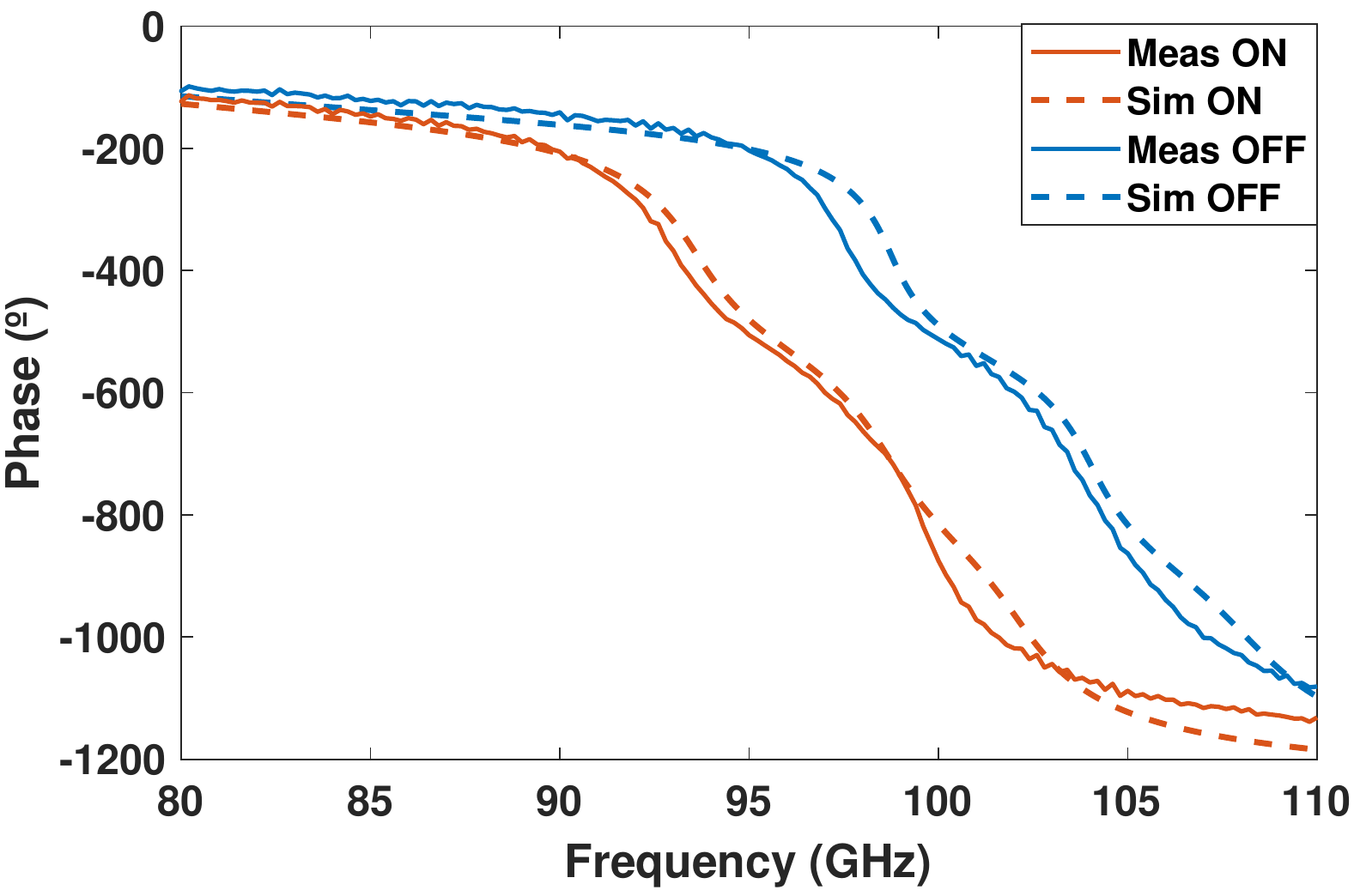}
    \vspace{-0.6cm}
    \caption{Phase of the reflection coefficient. Dashed lines indicate simulation data and flat lines indicate experimental measurements.}
    \label{fig:phase_spectrum}
    \vspace{-0.5cm}
\end{figure}
Once the different tilt angles along $z$ are obtained for each timestamp of a certain transition, and therefore $\hat{n}(z,t)$ is known, it is necessary to compute the permittivity tensors in accordance with the formulation of section \ref{sec:Permittivity Tensor Calculation}.
After the permittivity model has been calculated using COMSOL, this is used in an electromagnetic simulator (CST Studio \cite{cst}) with the aim of obtaining the electrical
parameters of the periodical reflective cell in RF.  

As sketched in Fig. \ref{fig:stratified}, in order to model the LC inhomogeneity along $z$ two different strategies have been used and compared, similarly to \cite{accurate}. On one hand, a stratified media has been considered, deploying different substrate layers within the LC. In this way, we consider the inhomogeneity along $z$ by partitioning the LC in N uniform layers in which the average tilt angle is assumed. With a large number of layers, this method is the most accurate but highly inefficient, even more than in \cite{accurate} due to the multiple simulations needed for a dynamics study. On the other hand, an average of the tilt angle across the entire cavity has been considered in order to work with a single-layer material encompassing the permittivity inhomogeneity across z. The purpose of doing so is to find a trade-off between accuracy and efficiency given that a multi-layer electromagnetic analysis is very costly. Note that this more precise than directly averaging the permittivity across $z$. This introduces a certain error to the computation, but greatly increases the efficiency of the simulation. Specifically, the latter method will become more accurate with extreme bias voltages, as the average and the local tilt values are almost the same across all $z$, and the difference with the stratified method will be negligible. This can be noticed in the top curves of Fig. \ref{fig:transition_up}, where the 150V excitation makes almost all molecules along $z$ to be rotated 90º at $T=35ms$, in contrast with the 15V excitation at $T=4s$. Therefore, in medium-voltage excitations, the stratified procedure will work slightly better but its complexity will increase abruptly. Fig. \ref{fig:images/layers comparison convergence all GT323001 75um 97GHz v3} compares the convergence of such stratified models by computing phase transitions in a 1, 3, 5 and 20-layered unit cell, showing that the single layer method is really precise, especially at large voltage excitations, as opposed to the analytical approximations. As will be seen later in the paper, these high-voltage excitations are of special interest for the overdrive technique to reduce reconfigurability times. The error made in the homogeneous case with respect to the stratified model is found to be reduced (around 20º in the worst case) and admissible, given that the corresponding degradation of the radiation pattern for such phase deviation is negligible \cite{quantization,encinar}. Moreover, it can be appreciated how for increasing voltages the difference between both methods vanishes (20º and 0.5º difference in 5V and 150V excitation respectively). The error is similar to that found in \cite{accurate} for statics, but it is generalized here for dynamics. Therefore, single-layer modelling is carried out for the following experiments with the aim of increasing computation efficiency at a negligible accuracy loss, especially at high voltages. However, the stratified approach could be followed for a perfectionist modelling at higher computational cost. For mid-range problems, increasing the number of layers until the accuracy converges is a reasonable procedure.

In order to validate the simulated dynamic results, the transient phase-curves at a certain frequency are compared against experimental data, captured from a reflectarray antenna whose unit cell is shown in Fig. \ref{fig:cavity}c, containing a LC cavity filled with GT3-23001 from Merck \cite{merck} ($k_{33}=34.5pN; k_{11}=24pN; \gamma_1=746mPas;  \varepsilon_{r||}=3.27;  \varepsilon_{r\perp}=2.47;  \Delta\varepsilon_q=4.6$) sandwiched below a Quartz superstrate ($\varepsilon_r = 3.78, tan \delta = 0.002$). The complete antenna consists on 60 x 60 identical cells (see Fig. \ref{fig:setup}a), and its phase response is shown in Fig. \ref{fig:phase_spectrum} for extreme excitations (OFF state corresponds to $V=0$ and ON state corresponds to $V>>Vth$) in stationary regime.

To experimentally acquire the dynamic cell phase curves at each timestamp for a specified frequency, incidence angle and bias signal, the setup shown in Fig. \ref{fig:setup}b and \ref{fig:setup}c has been implemented. An arbitrary waveform generator (Keysight 33611A), programmed to output different bias signals, drives the LC of the reflectarray antenna through a x15 voltage multiplier. In order to ensure a specular reflection, all unit cells of the antenna are short-circuited so that the LC is excited equally along the entire array. The waveform generator sends a SYNC signal to a vector network analyzer (VNA) to guarantee a timely capture of the transition. The VNA has been previously calibrated with a metallic plane reference and equipped with a pair of horn antennas, and captures the evolution of the transmission scattering parameters at 97 GHz and 102 GHz when the bias signal sequence starts. Those frequencies were selected since they show maximum phase range within the band of operation of the 360º cell, thus allowing the detection of the maximum phase errors. Both the experiments and the electromagnetic simulations have been carried out considering an impinging angle of 30º with respect to the normal of the reflectarray plane ($\phi_i=0$º, $\theta_i=30$º).

\begin{figure*}[!htb]
    \centering    
    \includegraphics[width=1\textwidth, height=6.5cm]{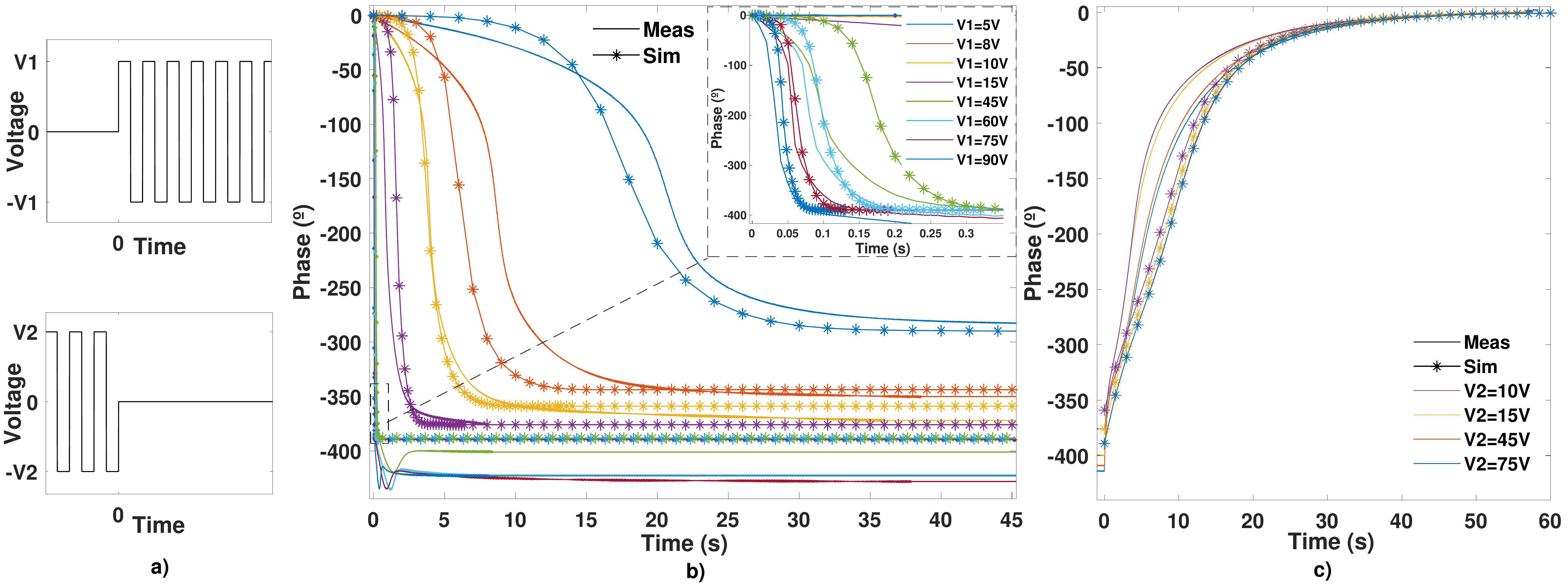}
    \vspace{-0.7cm}
    \caption{GT3-23001 phase transition dynamics at 102 GHz. a) 1 kHz biasing signal for excitation (top) and relaxation (bottom) dynamics b) Excitation transient phase for different V1 values c) Relaxation transient phase for different V2 values. The asterisk marker indicates simulation data and flat lines indicate experimental measurements.}
    \label{fig:transition_modelo}
\end{figure*}

Fig. \ref{fig:transition_modelo}b compares simulations and measurements of the transient reflected field phase at 102 GHz as a function of time for different voltage rise transitions starting from idle, verifying that the tool can predict relatively close the actual cell behaviour. The applied 1 kHz square signal allows to assume an amplitude equal to its RMS value in Equation (\ref{eq:ericksenleslie2}).
The model has been validated with relaxation measurements as well. Fig. \ref{fig:transition_modelo}c shows simulation and measurement data of the reflected field phase evolution at 102 GHz for different decay transitions parting from varying voltages. The corresponding biasing signals are shown in Fig. \ref{fig:transition_modelo}a.

As can be seen, the temporal model of the LC transitions matches the experimental data both in rise and decay time, as well as in phase range. It can be noticed how, as expected, while the rise transitions are highly dependent on the applied voltage, the decay transitions are quite similar regardless of the driving amplitude at $T=0$. Small discrepancies ($<$30º) between the expected and measured phase range in permanent regime can be explained by the phase curve in Fig. \ref{fig:phase_spectrum}, where the phase difference between both states slightly differs. It should be mentioned that this range of error becomes negligible when conforming a full radiation pattern, as it is equivalent to a 3-bit phase quantization, which generally suffices to synthesize a collimated beam and only deteriorates gain by 0.2 dB and SLL by 0.8 dB \cite{phd,quantization}. Errors in transient regime (up to 200º in the worst case of $V1=8V$) can be associated to different sources, including the 1-layer assumption and especially the pre-tilt angle estimation and LC RF characterization. This error could be minimized by choosing a denser layer stratification and assuming a lower computational efficiency, considering their trade-off. Additionally, the different error sources could be compensated in the model by performing, a posteriori, an effective parameter tuning for each voltage curve using measurement data.
\begin{comment}
\begin{figure}[!htb]
    \centering
    \includegraphics[width=1\columnwidth]{images/Comparison_Dynamic_LC_behaviour_0toX_GT323001_pretilt2_75um_102GHz_v8.pdf}
    \vspace{-0.7cm}
    \caption{GT3-23001 phase transition dynamics at 102 GHz when excited by a 1 kHz square signal with different amplitudes. The asterisk marker indicates simulation data and flat lines indicate experimental measurements.}
    \label{fig:transition_subida}
\end{figure}

\begin{figure}[!htb]
    \centering
    \vspace{-0.3cm}
    \includegraphics[width=1\columnwidth]{images/Comparison_Dynamic_LC_behaviour_Xto0_V_GT323001_75um_102GHz_v6.pdf}
    \vspace{-0.8cm}
    \caption{GT3-23001 phase transition dynamics at 102 GHz when relaxed to 0V from a 1 kHz square signal with different amplitudes. The asterisk marker indicates simulation data and flat lines indicate experimental measurements.}
    \label{fig:transition_bajada}
\end{figure}
\end{comment}

\section{Biasing Synthesis Techniques}

The dynamic modelling of LC multi-resonant cells, obtained and experimentally validated in Section \ref{sec:Model Results and Experimental Validation}, enables the development of bias voltage design techniques through simulations, which allow an improvement on the antenna reconfigurability times. These methods can be used to reduce both the relaxation times (underdrive) and the rising times (overdrive) although, as will be seen next, the main improvement occurs in the rising transitions.
\begin{figure}[!htb]
    \centering
    \vspace{-0.3cm}
    \includegraphics[width=1\columnwidth]{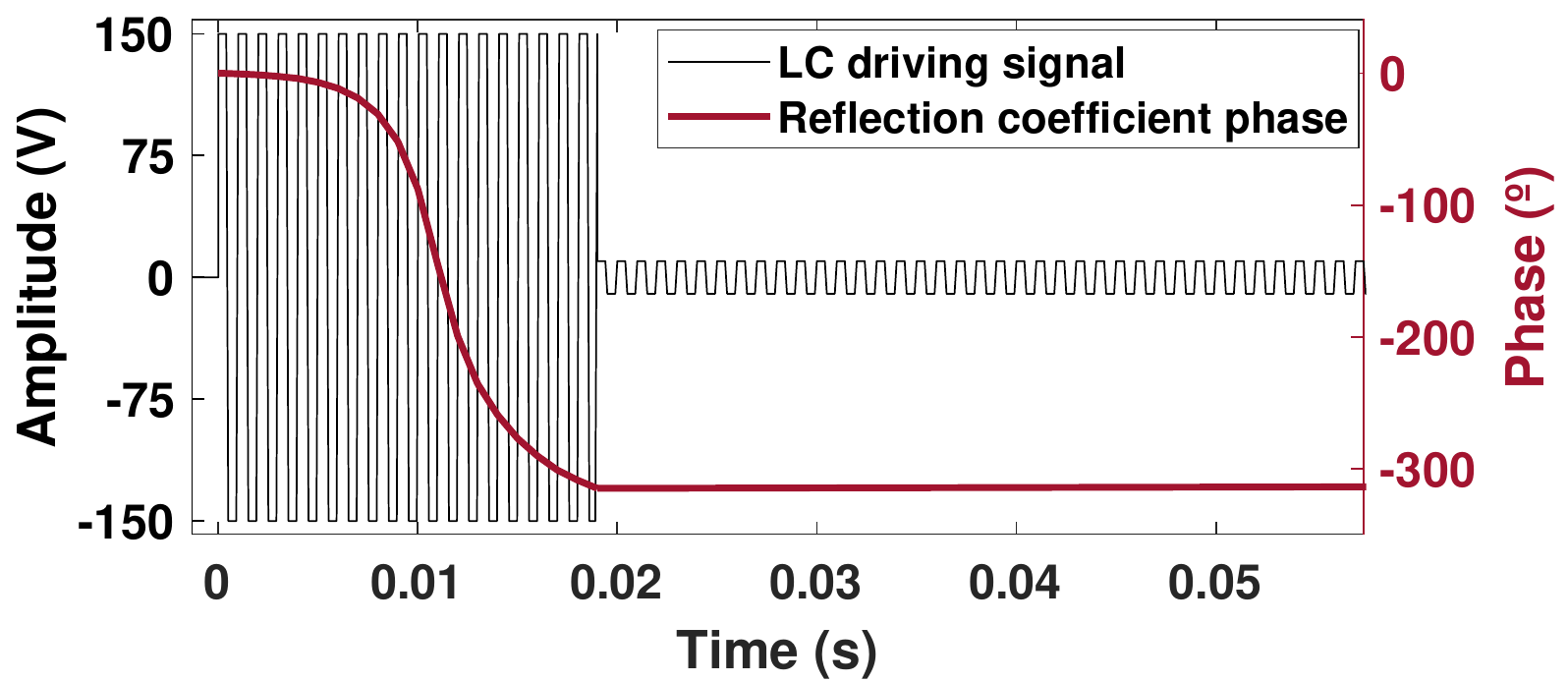}
    \includegraphics[width=1\columnwidth]{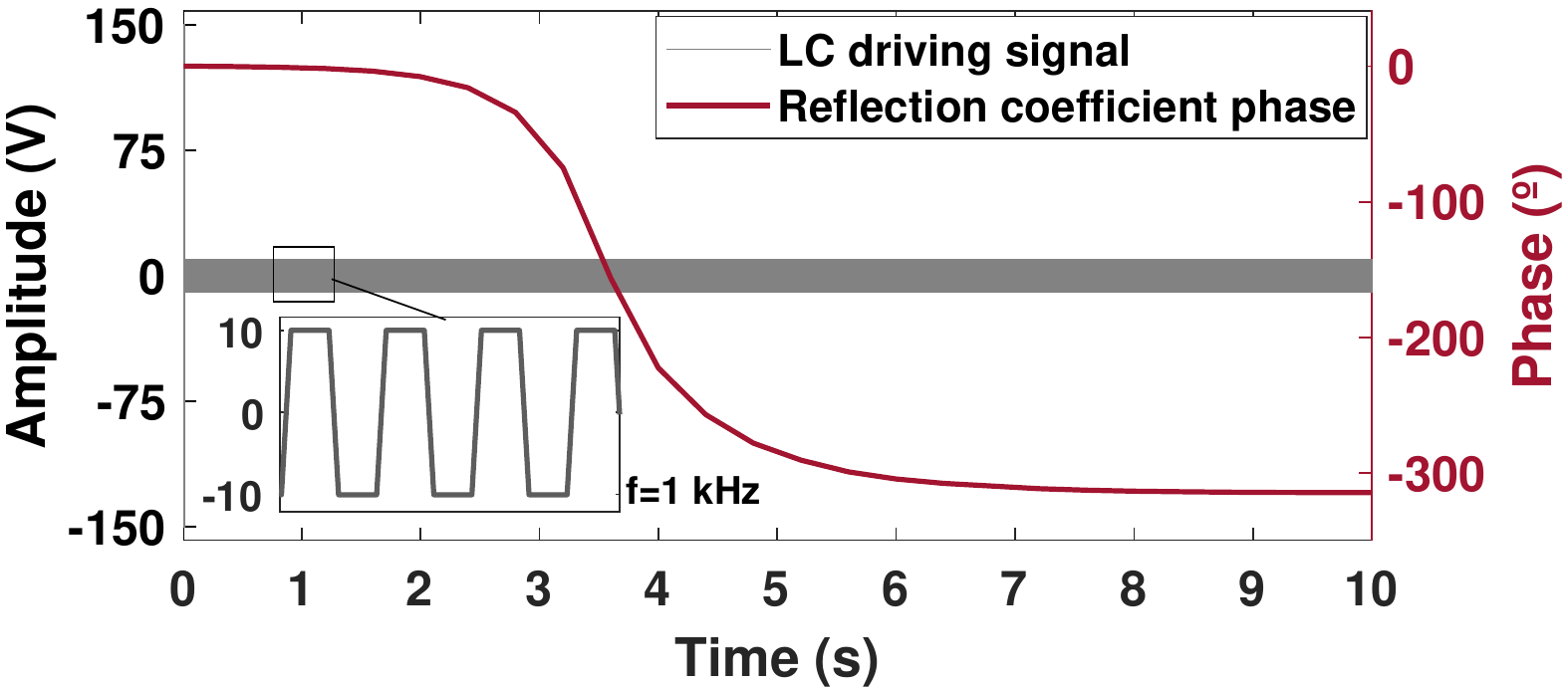}
    \vspace{-0.7cm}
    \caption{Overdrive technique (top) and nominal biasing (bottom) close-up look.}
    \label{fig:overdrive_overview}
    \vspace{-0.3cm}
\end{figure}

Specifically, by leveraging an accurate temporal control of the LC, a timely driving of the LC cells can be used to accelerate the transition times between phase states. That is, by overdriving it in the rise transitions, the LC orientation time can be accelerated when the electric field is increased (i.e. rotating the molecules towards parallel to $z$). This is achieved by using during a short period of time a larger voltage than the nominal biasing voltage (i.e. the voltage in which the cell presents the desired phase shift in permanent regime, after the molecules stopped rotating). This can be seen in Fig. \ref{fig:overdrive_overview}, where the overdrive LC driving signal amplitude is 150V until the 10V objective reflection coefficient phase is achieved, instant in which we switch the excitation to the nominal 10V signal amplitude. Similarly, by underdriving it (sometimes referred to as undershoot), the LC orientation time can be accelerated when the electric field is decreased (i.e. rotating the molecules towards perpendicular to $z$). In this case, the LC is briefly driven at a lower voltage than the nominal one. Additionally, by using dual frequency LC, the overdrive technique can be used to accelerate both transitions.
The overdrive technique has been used in optic devices in the past \cite{overdrive1}, but by using approximations instead of an accurate profiling, and not in the mm-wave regime where the cell thickness and modelling become problematic.

In the case of aperture antennas where the objective parameter is the pixel phase, the design procedure of the LC overdriving signal for quickly achieving the desired phase of an array cell, defined by its dimensions and incidence angle, is the following:

\begin{enumerate}
    \item Identify the nominal voltage that achieves the desired phase shift in stationary state.
    \item Compute $\theta(z,t)$ for the rise transition towards the nominal voltage, by solving Equation (\ref{eq:ericksenleslie2}).
    \item Find the phase-time curve of such transition by solving the structure electromagnetically, for each timestamp, after finding $\overline{\overline{\varepsilon_r}}$ from Equation (\ref{eq:eptensor}). 
    \item Repeat steps 2 and 3 for the rise transition towards the maximum voltage.
    \item Pick, from the rise transition towards the maximum voltage, the timestamp in which the instantaneous phase matches the converged phase of the nominal transition.
\end{enumerate}

\begin{figure}[t]
    \centering
    \vspace{-0.1cm}
    \includegraphics[width=1\columnwidth]{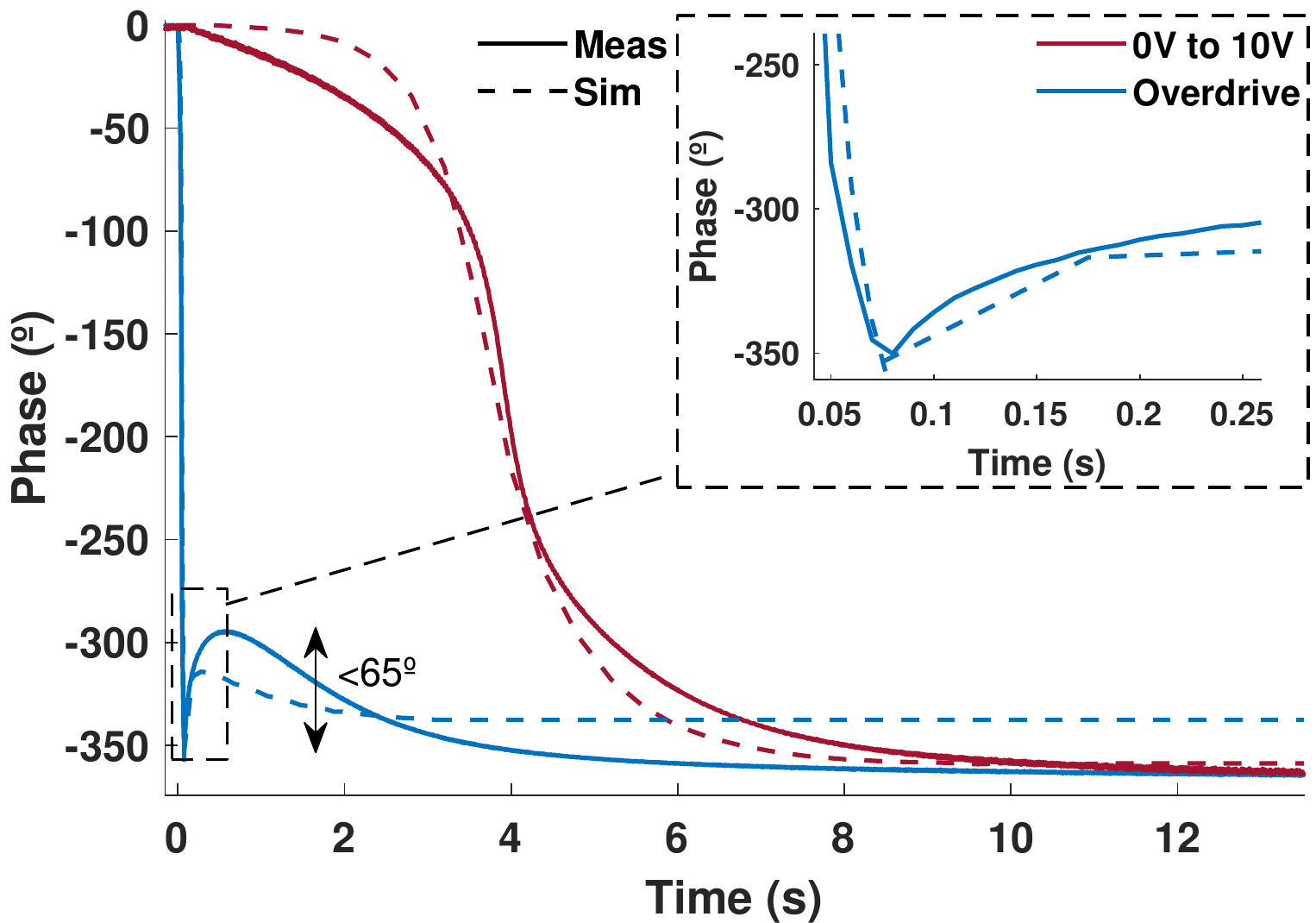}
    \vspace{-0.8cm}
    \caption{Measured and simulated phase of reflection coefficient at 102 GHz during a 0V to 10V transition. To quickly achieve the 10V state phase, we overdrive the LC to 75V during 75ms and then switch to the nominal 10V excitation.}
    \label{fig:overdrive 10V}
    \vspace{-0.3cm}
\end{figure}
\begin{figure*}[t]
    \vspace{-0.4cm}
    \centering
    \includegraphics[width=1\textwidth]{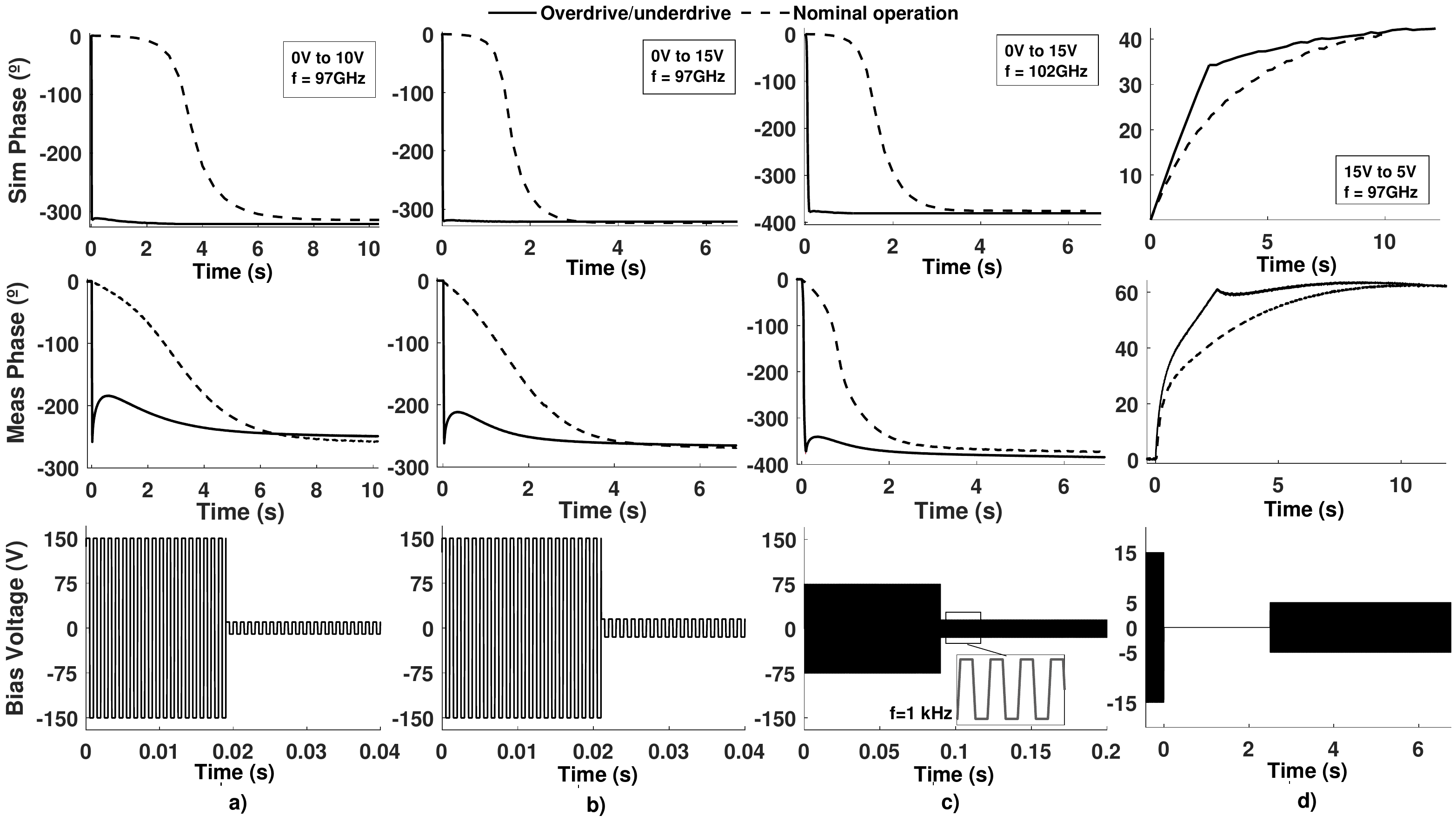}
    \vspace{-0.8cm}
    \caption{Phase transition between states using overdrive/underdrive and nominal excitations. Top row shows Simulations and middle row shows Measurements of a) 0V to 10V at 97 GHz, using a 150V overdrive for 19ms; b) 0V to 15V at 97 GHz, using a 150V overdrive for 21ms; c) 0V to 15V at 102 GHz, using a 75V overdrive for 90ms; d) 15V to 5V at 97 GHz, using 0V underdrive for 2.5s. Bottom row shows the applied overdrive/underdrive bias signal.}
    \label{fig:overdrive_underdrive_simsymeas}
    \vspace{-0.5cm}
\end{figure*}

\begin{table*}[h]
    \centering
    \begin{threeparttable}[b]
    \caption{Electrical beam-steering works comparison at both cell and antenna level}
    \begin{tabular}{|c|c|c|c|c|c|c|c|c|}
        \hline
        \textbf{Work} & \textbf{Technology} & \textbf{Freq. (GHz)} &  \textbf{Losses} & \textbf{PR} & \textbf{Ton} & \textbf{Toff} & \textbf{Pol.} &  \textbf{Complete antenna parameters}\\
        \hline
        \cite{adualpol} & PIN & 7.45 & 12dB & 180º & $<$1ms\tnote{*} & $<$1ms\tnote{*} & DL & G: 21dB. SA: $\pm$40º 2D. BW: 0.85 (3dB)\\
        \hline
        \cite{dualband} & Varactor & 11.3/14.7 & 2/3dB & 300º & $<$1ms\tnote{*} & $<$1ms\tnote{*} & C & G: 14/16.6dB. SA: $\pm$24º 2D. BW: 0.85/0.63 (3dB)\\
        \hline
        \cite{realization} & Varactor & 23.5 & 3dB & 320º & $<$1ms\tnote{*} & $<$1ms\tnote{*} & SL & SA: $\pm$60º 1D\\
        \hline
        \cite{vo2} & VO$_2$ & 32 & 1dB & 300º & 12ms & 2s & SL & -\\
        \hline
        \cite{mems} & MEMS & 11.2 &  0.5dB & 180º & $<$1ms\tnote{*} & $<$1ms\tnote{*} & DL & -\\
        \hline
        \cite{reflectarray2} & LC & 24.1 & 4dB & 360º & 5s\tnote{*} & 10s\tnote{*} & SL & G: 20.2dB. SA: $\pm$45º 2D. BW: 4 (3dB)\\
        \hline
        \cite{folded} & LC & 78 & 12dB & 270º & 5s\tnote{*} & 10s\tnote{*} & SL & G: 25.1dB. SA: 17º 1D. BW: 3 (1dB)\\
        \hline
        \cite{lcreflectarray}** & LC & 100 & 6dB & 360º & 5s & 10s & SL & G: 19.4dB. SA: 55º 1D. BW: 6 (3dB)\\
        \hline
        This work** & LC + overdrive & 100 & 6dB & 360º & $<$50ms & 5s & SL & -\\
        \hline
    \end{tabular}
    \vspace{0.1cm}
    \label{tab:works}
    \begin{tablenotes}
        \item[*] Estimated considering technology. ** Same cell design and LC material employed to facilitate the temporal comparison.\\ PR = Phase range; DL = Dual linear; SL = Single linear; C = Circular; G = Gain; SA = Scan angle; BW = Gain bandwidth.
    \end{tablenotes}
    \end{threeparttable}
    \vspace{-0.4cm}
\end{table*}

Then, the driving signal consists on modifying the amplitude of the nominal biasing signal to the maximum voltage between $t=0$ and the obtained timestamp. Regarding the underdriving signal design, the procedure is dual by using a drop transition towards a zero voltage. Given that implementing these strategies properly requires a very precise knowledge of the LC dynamics in the cells, and given that it is unfeasible to obtain such curves for each cell and incidence angle, it is necessary an accurate enough model like the one proposed in this work that provides such information through simulations.
%\begin{figure}[!htb]

In order to validate such technique, different temporal driving signals have been computed so as to reduce the switching times for different state transitions, and experimental measures have been obtained using those excitations. %as shown in Fig. \ref{fig:overdrive 10V} and Fig. \ref{fig:overdrive_underdrive_simsymeas}.
In Fig. \ref{fig:overdrive 10V}, a comparison between simulated and measured data for both overdrive and nominal excitations is shown for a 0V to 10V transition. As can be seen, the overdrive technique applied to the reflectarray antenna allows to accelerate the rise time by a factor 100X.
In Fig. \ref{fig:overdrive_underdrive_simsymeas}, measurements and simulations for both overdrive and underdrive techniques are compared against a normal operation for different phase transitions at two representative frequencies in the band of design, and the corresponding bias signals are shown. As can be observed, the overdrive strategy allows for a much quicker phase drop than the normal operation, reducing in several orders of magnitude the switching times. Although the underdrive technique also shows some time reduction (Fig. \ref{fig:overdrive_underdrive_simsymeas}d), its effect is not so pronounced, as the decay transition is less dependent on the excitation. Moreover, since the rise transitions are completely dependent on the driving amplitude, the voltage can theoretically be further increased to reduce the rising times (in practice, we will be restricted by equipment and the limited cavity impedance creating a short-circuit), but it can not be underdriven beyond 0V since it is the absolute magnitude of $E$ what makes LC molecules to rotate. Notwithstanding, the decay times can be further reduced by choosing a less viscous LC, or by employing dual-frequency LC, which can in turn benefit from this overdrive technique to decrease relaxation times. Additionally, both excitation and relaxation times may be drastically decremented by properly combining the overdrive techniques and the previously mentioned polymerizable materials, although future work on its characterization is required. It should be mentioned that even though the voltage is increased significantly during these transitions, the overall power consumption is minimum, as the LC cell hardly consumes any current.
%the introduced model could allow an effective characterization of polymerizable LCs.

An interesting phenomenon that occurs in some of the captured phase transitions is a significant rebound of the phase right after the objective phase is initially achieved in the overdrive excitation, as shown in Fig. \ref{fig:overdrive 10V} and the middle row of Fig. \ref{fig:overdrive_underdrive_simsymeas}. This bounce is the manifestation of both the bias signal commutation (which causes the molecular reorientation of Fig. \ref{fig:transition_up}c) and the backflow effect \cite{highspeed,pretilt, kickback,bck}, which appears as a consequence of working in the high-voltage regime in thick cells and that our model did not completely capture. This is a well known phenomenon that, if one has access to the LC Leslie coefficients, either through manufacturer data or through experimental estimation \cite{backflow,lesliecoef}, could be included in the problem to make the model more accurate but more computationally costly. Even though this effect deteriorates the experimental measurements as the phase oscillates slightly ($<$65º in Fig. \ref{fig:overdrive 10V}) until reaching the final value, we can still approach the objective phase state much faster than under a nominal excitation, being the instantaneous phase during such transient effect $\pm$20\% deviated only.

Overall, the predicted and measured reflectarray cells transitioned a maximum of 250X and an average of 100X faster between phase shift states when using overdrive techniques, as compared to using nominal excitations. On the other hand, by completely removing the biasing voltage temporarily, the underdrive excitations shortened in average a 2X time factor to achieve 90\% of the objective phase, as compared to the nominal excitations.
To put this into perspective, Table \ref{tab:works} compares different works on electrical beam-steering phase-shift LC metasurfaces and unit cells, and its performance including transition times. Additionally, a comparison with other reconfigurable metasurface technologies is included. 
\vspace{-0.2cm}
\section{Conclusion}

LC-based reconfigurable metasurfaces are promising candidates for developing electrically large aperture antennas supporting the future generation of communications, given their easiness of manufacturing, low cost and wide operating frequency ranges. However, switching times between phase states must be reduced before they can be widely used in real-time applications. In this work, a dynamical model of LC transitions for different excitations beyond the known approximations is presented and validated in order to achieve a temporal control of the unit cell phase, useful for both reflective and transmissive multi-resonant cells. A further analysis on the LC stratified model is also provided by considering a different number of simulated layers, concluding that a reduced number of layers (N=20) is needed in the worst case, although using an effective tensor (N=1) will be enough to achieve reasonable accuracy most of the times, and useful to perform efficient electromagnetic simulations. Even though the effect of the different LC driving excitations on the phase change can be carried out through both measurements and simulations, a generalization in frequency, incident angle, cell designs and LC materials could be cumbersome to do by means of measures. Instead, a simulation tool like the one introduced in this work allows for a fast and accurate estimation of control signals to introduce the temporal parameter in the design space of electrically large antennas. In turn, this allowed to design and validate an overdriving technique capable of drastically reducing transition times by one or two orders of magnitude in a simple way. In the rising case, reductions are of a factor 100 while in the relaxation case improvements are less drastic. However, this strategy can be used together with others to enhance the whole LC dynamic behaviour and obtain reduced antenna scanning times.

\vspace{-0.3cm}

\small{
\bibliographystyle{ieeetr}
\bibliography{ref}
}

\end{document}